\documentclass[useAMS, usenatbib, a4paper]{mn2e}

\usepackage{aas_macros}	
\usepackage{amsmath}		
\usepackage{deluxetable}	
\usepackage{lscape}		
\usepackage{graphicx, color}	

\def\fsc{\alpha_f} 

{\catcode`\@=11                                                  
   \gdef\SchlangeUnter#1#2{\lower2pt\vbox{\baselineskip 0pt\lineskip0pt    
   \ialign{$\m@th#1\hfil##\hfil$\crcr#2\crcr\sim\crcr}}}}

\title[Magnetar atmosphere models]{Models of hydrostatic magnetar atmospheres at high luminosities}
\author[T. van Putten et al.]{T. van Putten,$^{1}$\thanks{E-mail:
T.vanPutten@uva.nl} A. L. Watts,$^{1}$ C. R. D'Angelo,$^{1}$ M. G. Baring$^{2}$ and C. Kouveliotou$^{3}$ \\
$^{1}$Astronomical Institute ``Anton Pannekoek,'' University of Amsterdam, Postbus 94249, 1090 GE Amsterdam, The Netherlands\\
$^{2}$Department of Physics and Astronomy, MS 108, Rice University, Houston, TX 77251, USA\\
$^{3}$Space Science Office, VP62, NASA/Marshall Space Flight Center, Huntsville, AL 35812, USA\\
}
\begin{document}

\date{Accepted 2013 June 17}

\pagerange{\pageref{firstpage}--\pageref{lastpage}} \pubyear{2013}

\maketitle

\label{firstpage}

\begin{abstract}
We investigate the possibility of Photospheric Radius Expansion (PRE) during magnetar bursts. Identification of PRE would enable a determination of the magnetic Eddington limit (which depends on field strength and neutron star mass and radius), and shed light on the burst mechanism.  To do this we model hydrostatic atmospheres in a strong radial magnetic field, determining both their maximum extent and photospheric temperatures.  We find that spatially-extended atmospheres cannot exist in such a field configuration:  typical maximum extent for magnetar-strength fields is $\sim$ 10m (as compared to 200 km in the non-magnetic case).  Achieving balance of gravitational and radiative forces over a large range of radii, which is critical to the existence of extended atmospheres, is rendered impossible in strong fields due to the dependence of opacities on temperature and field strength.   We conclude that high luminosity bursts in magnetars do not lead to expansion and cooling of the photosphere, as in the non-magnetic case.  We also find the maximum luminosity that can propagate through a hydrostatic magnetar atmosphere to be lower than previous estimates.   The proximity and small extent of the photospheres associated with the two different polarization modes also calls into question the interpretation of two blackbody fits to magnetar burst spectra as being due to extended photospheres.
\end{abstract}


\section{Introduction}

Photospheric Radius Expansion (PRE) events can occur during bursts on neutron stars when the luminosity of the object reaches the Eddington Luminosity, i.e. where the radiation force balances the gravitational one:
\begin{equation}
\label{eq:Ledd}
L_{\rm Edd} \equiv \frac{4\pi GM_*c}{\kappa_\rmn{Th}},
\end{equation}
(where $M_*$ is the stellar mass and $\kappa_\rmn{Th}$ the Thomson scattering opacity) and the large radiation pressure forces the atmosphere to expand outwards. For the hydrogen atmosphere of a $1.4 M_\odot$ neutron star, $L_{\rm Edd} = 1.8\times 10^{38} \rm{erg}\rm{s}^{-1}$, while for a helium atmosphere it is twice that value. As a result, the photosphere moves to a much larger radius, corresponding to a drop in temperature $T$. For a neutron star with a modest magnetic field (up to $\sim10^{12}~\rm{G}$), Compton scattering dominates the opacity in the atmosphere and various relativistic effects allow the atmosphere to expand up to hundred kilometres, so that the temperature of the expanded photosphere drops out of the X-ray range altogether \citep{Hoffman1978, Paczynski1986}. The hallmark of PRE in neutron stars is thus a `double-peaked' structure in the X-ray light curve of a burst, in which the flux increases to a maximum and then drops sharply (indicating the black-body temperature has decreased), before rising again steeply to a slightly larger maximum as the bolometric luminosity drops again and the photosphere contracts \citep{Paczynski1983}.

PRE is characteristically seen in Type I X-ray bursts from accreting neutron stars, in which the build-up of accreted material leads to a thermonuclear explosion on the surface of the star, causing a huge increase in luminosity. PRE bursts have typically been used to constrain the mass and radius of the neutron star, thus potentially constraining the equation of state of dense matter \citep[e.g.][]{Damen1990, Galloway2003a, Ozel2009a, Ozel2010a, Steiner2010a, Suleimanov2011a}. However, PRE is generically driven by high luminosities, irrespective of the underlying energy source. This has led to the recent suggestion that it might also happen in bright bursts from magnetars \citep{Watts2010} -- isolated neutron stars with dipole magnetic fields above $\sim10^{13}~\rm{G}$ -- whose bursts (which occur over a wide range of luminosities) are thought to be powered by large-scale reconfiguration of the decaying magnetic field \citep{Thompson1995}. \cite{Watts2010} argued that observing PRE in magnetar bursts could put interesting constraints on the emission mechanism, magnetic field strength and mass-radius relationship for magnetars.

The suggestion that PRE might happen in magnetar bursts was motivated by the August 2008 observation of a large ($L_X \sim 7\times 10^{39}$ erg s$^{-1}$) burst from SGR 0501+4516, which showed a double-peaked light curve similar to those seen in Type I X-ray bursts. In their paper on this burst, \cite{Watts2010} laid out several criteria required for PRE to occur, and argued that they were in general met for magnetar bursts. For PRE to occur in a neutron star, the flux must be emitted from an optically thick region, the radiation pressure must be sufficient to overcome gravity and other confining forces, the emitting region must remain optically thick during the expansion (so that the emission remains close to blackbody and effective temperature decreases with increasing photosphere radius), and the opacity must increase with distance from the star. The last point is slightly subtle: in order for the photosphere to expand, the luminosity $L$ must remain close to the critical luminosity ($L_{\rm cr}$)  needed to balance radiation pressure with the confining forces\footnote{Here and throughout the paper, we follow \cite{Paczynski1986} in defining $L_{\rm cr}$ as the actual maximum luminosity as a function of radius, modified by the changing opacity and gravitational redshift, whereas $L_{\rm Edd}$ is strictly given by eq \ref{eq:Ledd}}. However, both these quantities are modified differently by the strong gravitational field, so that $L/L_{\rm cr} \propto 1+z$, where $z$ is the gravitational redshift. To ensure that this quantity does not decrease with radius, which would make expansion impossible \citep{Paczynski1986}, the opacity (which determines $L_{\rm cr}$) must therefore increase with radius. In Type I X-ray bursts, this is effected by the Klein-Nishina corrections which reduce the Thomson cross-section at high temperatures close to the stellar surface.

The presence of a magnetar-strength magnetic field complicates the hydrostatic expansion of the atmosphere in three significant ways. In the  `trapped fireball' picture of magnetar bursts \citep{Thompson1995}, the huge release of magnetic energy leads to the creation of a pair-plasma, so that the atmosphere is dominated by this dense pair-gas, rather than baryonic matter ablated from the star's surface. Further, closed field lines provide a strong confining force for both baryonic and leptonic matter, since plasma cannot easily move perpendicular to the field. A straightforward calculation (see section \ref{sec:Lcr}) demonstrates that a very strong field can easily confine even the largest giant flares with $L\sim10^{44} \rm{erg}; \rm{s}^{-1}$ \citep{Lamb1982}, so that PRE in a magnetar will likely only occur in open field line regions. 

The final effect of the magnetic field, and the most important one in the present work,  is to modify the electron scattering cross-sections by several orders of magnitude, depending on the polarization state of the scattering photon. The strong magnetic field suppresses electron motion perpendicular to the field, so that photons that try to excite this motion (i.e. that are polarized perpendicular to B; the `Extraordinary' or E-mode) have a greatly reduced scattering cross-section compared to the Thomson scattering cross section $\sigma_\rmn{Th}$, while photons polarized parallel to B (the `Ordinary', O-mode) are largely unaffected. The modified polarization-dependent cross-sections will increase the critical luminosity (sometimes called the `magnetic Eddington limit') by several orders of magnitude \citep{Thompson1995, Miller1995}. Additionally, since the E-mode cross-section scales roughly as  $(T/B)^{2}$ (see Equation \ref{eq:crosssec_red}), the opacity increases steeply with distance from the star from the decrease in field strength, and becomes strongly temperature dependent. $L_{\rm cr}$ is therefore a strong function of radius. As we will demonstrate, the strong temperature and field dependence of the opacity has a profound effect on the structure of the magnetar atmosphere in comparison to the non-magnetic case.

The object of this paper is to explore the structure of a magnetar atmosphere at very high luminosities. We follow the approach of \cite{Paczynski1986} and calculate the structure of a series of hydrostatic atmospheres with different masses, base temperatures, and magnetic field strengths, solving the equations for stellar structure. The main difference from \cite{Paczynski1986} is in our consideration of the opacity, which is dominated by electron scattering in both cases. They consider a non-magnetic star, for which the electron scattering cross-section is the Thomson one (modified at high temperatures by Klein-Nishina corrections). We instead use the cross-sections modified by the strong magnetic field, so that the radiation is split into two polarization modes, and only E-mode photons diffuse through the atmosphere.

The paper runs as follows. In Sections \ref{sec:Lcr}-\ref{sec:structure} we discuss the concept of a `critical luminosity' in more depth, and present the equations we use to calculate the structure of the atmosphere. Section \ref{sec:opacity} focuses on the electron scattering cross-sections in a super-strong magnetic field. The field introduces several complications (such as the dependence on photon energy of the scattering cross-section and the presence of the cyclotron resonance), and we explain how we calculate the effective opacity for a thermal photon distribution. Section \ref{sec:results} presents the main results of our calculations, and Section \ref{sec:comparison} compares our results to previous calculations of the critical luminosity. Finally, in Section \ref{sec:extensions}  we examine additional physical processes (most significantly, a pair plasma gas and magnetic confinement from closed field lines), and argue that neither of these are likely to affect the qualitative conclusions of the paper, by which we mean the non-existence of hydrostatic extended magnetar atmospheres.

\section{Model}
\label{sec:model}
The structure of the atmosphere of a magnetar can be calculated from
the equations of stellar structure.  We do this from the surface
of the magnetar to a point far away from the star where the atmosphere can be said to have ended.
We describe the various equations in detail, starting with the critical luminosity and magnetic field in Section \ref{sec:Lcr}
and the stellar structure equations in Section \ref{sec:structure}. The form of the opacity in a strong
magnetic field is discussed in Section \ref{sec:opacity}. We then treat the boundary conditions and computational
method in Section \ref{sec:method}, and briefly summarize our model.

\subsection{Critical luminosity in a super-strong magnetic field}
\label{sec:Lcr}

Our goal is to compute the ratio between the local luminosity $L$ and the critical luminosity
$L_\rmn{cr}$ throughout the atmosphere, with the latter defined as the luminosity at
which the outward radiation force is exactly balanced by inward forces.
If $L/L_\rmn{cr}$ is greater than one at any point, the atmosphere will be
unstable there, and thus not in hydrostatic equilibrium.

In this work we define $L_\rmn{cr}$ to be the critical luminosity at which the radiative force exactly matches
gravity, reserving the term Eddington luminosity for the special case where the opacity is given by the Thomson
scattering opacity.
This critical luminosity is given by
\begin{equation}
 \label{eq:Lcr}
 L_\rmn{cr}=\frac{4\pi cGM}{\kappa}\left(1-\frac{R_\rmn{g}}{r}\right)^{-1/2},
\end{equation}
where $R_\rmn{g}=2GM/c^2$ is the gravitational radius, $M$ is the mass of the star, $\kappa$ is opacity,
$r$ is the radius measured from the centre of the star
and $c$ and $G$ are the usual constants. 

In a magnetar atmosphere the opacity will be reduced by
several orders of magnitude from the Thomson opacity \citep[][]{Thompson1995, Miller1995}, giving a critical luminosity
of order $10^{40}-10^{41}$ erg s$^{-1}$, which is
in the range of typical luminosities of magnetar bursts, and
several orders of magnitude larger than the Eddington luminosity
of $2\times 10^{38}$ erg s$^{-1}$. This critical luminosity is inversely proportional
to the opacity, which means that in a super-strong magnetic field it will be dependent on the magnetic field strength,
and thus be variable throughout the atmosphere.
We discuss the effect of the magnetic field on the opacity in detail in Section \ref{sec:opacity}.

In closed field line regions, gravity will not be the dominant confining force. Here,
plasma is trapped by suppressing the transport
of charges across field lines, and so
the radiation force is counteracted by magnetic stress.  In a magnetar this effect
will dominate gravity, giving for the critical trapping luminosity \citep{Lamb1982}
\begin{equation}
 \label{eq:intro:Lcr_mag}
 L_{\rmn{tr}}\simeq 2.1\times 10^{49}\left(\frac{B}{B_\rmn{cr}}\right)^{2}
  \left(\frac{R_\star}{10\,\rmn{km}}\right)^{2} \rmn{erg}\,\rmn{s}^{-1},
 \end{equation}
where $B$ is the magnetic field strength,
$R_\star$ the radius of the neutron star and
$B_\text{cr}=4.4\times 10^{13}$ G is the quantum critical magnetic field strength,
the field strength for which the cyclotron and rest mass energies of the electron are equal.

For a typical magnetar this critical luminosity will be above $10^{50}$ erg s$^{-1}$,
far more than the luminosity of even the rare giant flares.
As PRE requires
the luminosity to reach the critical luminosity to cause expansion, it is clear that PRE can only ever
occur in open field line regions.  In the remainder of this work we will therefore only treat open field line regions.

Treating only open field lines does not limit the applicability of our models greatly. Even in the trapped fireball
model \citep{Thompson1995, Thompson2001a}, where the radiation is emitted from the surface of
a fireball trapped in the closed magnetic field lines, most of the radiation will be emitted through open field-line regions.
This is because the fireball is the hottest and thus most luminous near its base, where the surface of the fireball
borders on the open field line regions, so that the radiation is emitted into the open field line region.

We consider a purely radial magnetic field with the field strength falling off as the field
of a dipole directly above either of the magnetic poles; the modifications incurred at 
non-zero colatitudes are not expected to qualitatively alter our conclusions.  We use the fully general relativistic equations for
the radial dependence of the magnetic field \citep[][]{Petterson1974, Wasserman1983a}

\begin{equation}
 \label{eq:B}
 B=-\frac{6\mu}{rR_\rmn{g}^2}\left[\frac{r}{R_\rmn{g}}\,\rmn{ln}\left(1-\frac{R_\rmn{g}}{r}\right)+\frac{R_\rmn{g}}{2r}+1\right],
\end{equation}
where $\mu$ is the magnetic dipole moment.

\subsection{Stellar structure equations}
\label{sec:structure}

We consider a spherically symmetric atmosphere model,
using the general relativistic equations of stellar structure given by
\citet{Thorne1977} to calculate the radial structure of a magnetar atmosphere.
The relevant equations are the equation of hydrostatic
equilibrium, the equation of energy transport, the equation for optical depth and the mass continuity equation.
We reformulate these equations as:

\begin{equation}
 \label{eq:dPdr}
 \frac{\rmn{d}P}{\rmn{d}r}=-\frac{GM\rho}{r^{2}}\left(1-\frac{R_\rmn{g}}{r}\right)^{-1}\left[1+
	 \frac{P+1.5P_\rmn{g}+4\sigma_\rmn{SB}T^4/c}{\rho c^{2}}\right],
\end{equation}

as a specialized form of the Oppenheimer-Volkoff equation,
\begin{equation}
 \label{eq:dTdr}
 \frac{\rmn{d}T}{\rmn{d}r}=\frac{T}{P}\frac{\rmn{d}P}{\rmn{d}r}\nabla,
\end{equation}
\begin{equation}
 \label{eq:dtaudr}
 \frac{\rmn{d}\tau}{\rmn{d}r}=-\kappa \rho \left(1-\frac{R_\rmn{g}}{r}\right)^{-1/2},
\end{equation}
\begin{equation}
 \label{eq:ddMdr}
 \frac{\rmn{d}\Delta M}{\rmn{d}r}=-4\pi r^{2}\rho \left(1-\frac{R_\rmn{g}}{r}\right)^{-1/2},
\end{equation}
where
$P$ is the total pressure,
$P_\rmn{g}$ is the gas pressure,
$T$ is the temperature,
$\rho$ is the density,
$\tau$ is the optical depth,
$\kappa$ is the opacity (Section \ref{sec:opacity}) and
$\Delta M$ is the rest mass of the atmosphere above radius $r$.
$\nabla$ is defined as d log $T/$ d log $P$, and depends on the manner in which energy is transported.
In principle, this is a combination of radiative transport, convective transport and conductive transport. However,
conduction is only significant compared to radiation at much higher densities than we consider \citep{Potekhin2007a},
while convection, which is included in the models from \citet{Paczynski1986}, is strongly suppressed by the magnetic field
in the magnetar case \citep{Rajagopal1996a}.
Thus, we set $\nabla=\nabla_\rmn{rad}$, which is given by

\begin{align}
 \label{eq:nabla_rad}
 \nabla_{\rmn{rad}}=&\left[\frac{\kappa L_{\infty}}{16\pi cGM(1-\beta)}
 \left(1-\frac{R_\rmn{g}}{r}\right)^{-1/2}+\frac{P}{\rho c^{2}}\right]\nonumber\\
 &\times\left[1+ \frac{P+1.5P_\rmn{g}+4\sigma_\rmn{SB}T^4/c}{\rho c^{2}}\right]^{-1},
\end{align}

with $L_\infty$ the luminosity as seen by an observer at infinity. Assuming blackbody emission, this luminosity
is linked to the temperature at the photosphere (where $r=R_\rmn{ph}$) 
through
\begin{equation}
   L_\infty \; =\; 4\pi\sigma_\rmn{SB}R_\rmn{ph}^2T_\rmn{ph}^4
   \left( 1- {\frac{ R_\rmn{g}}{R_\rmn{ph}}} \right)\quad ,
 \label{eq:Stephan_Boltzmann_GR}
\end{equation}
a Stefan-Boltzmann law modified by general relativity, which imposes an 
effective reduction in solid angles by the factor $1-R_\rmn{g}/R_\rmn{ph}$. Here $T_\rmn{ph}$ is the temperature
at $r=R_\rmn{ph}$ and
$\sigma_\rmn{SB}$ the Stefan-Boltzmann constant.

Equations (\ref{eq:dPdr}-\ref{eq:Stephan_Boltzmann_GR}) are derived from \citet{Thorne1977} by dropping the gravitational acceleration correction factor $4\pi r^3 P / Mc^2$, which is always less than $10^{-10}$ in our models, and using the `total mass' from \citeauthor{Thorne1977} as our mass. These equations reduce to those given by \citet{Paczynski1986} when combined with Equation \eqref{eq:pressure}, correcting two typographical errors in Equations (3a) and (4b) of that work.

We consider a purely radial magnetic field and use an ideal gas equation of state, so that
the pressure in the atmosphere is simply given by the sum of radiation pressure and gas pressure:
\begin{equation}
 \label{eq:pressure}
 P=P_\rmn{g}+P_\rmn{r}, \quad P_\rmn{g}=\frac{k\,T\,\rho}{\mu\,m_\rmn{H}},
 \quad P_\rmn{r}=\frac{4\sigma_\rmn{SB}}{3c} T^4, 
\end{equation}
where $T$ is the temperature, $\rho$ the density, $\mu$ the mean molecular
mass per free particle (both ions and electrons), $k$ the Boltzmann constant and $m_\rmn{H}$
the mass of a hydrogen atom. 
This equation for the radiation pressure assumes the gas is in
thermodynamic equilibrium with the radiation field, an assumption we will revisit in Section \ref{sec:nonlte}.
We ignore the effects of any mildly relativistic electrons, as these will not change our qualitative results.

We assume the atmosphere to be entirely composed of fully ionized helium.
This is not necessarily a realistic composition
for a magnetar atmosphere, but does make for easy comparison to the nonmagnetic case.
We will briefly discuss the effect of a more realistic composition, including an electron positron pair plasma,
in Section \ref{sec:composition}.

\subsection{Opacity}
\label{sec:opacity}

The opacity in a super-strong magnetic field is strongly reduced from the Thomson opacity, because
the scattering cross section of one of the two polarization modes of the photons is suppressed by the
magnetic field. Therefore, the opacity decreases strongly with increasing magnetic field strength,
which means that the critical luminosity increases with magnetic field strength. We will now go into
what this means for the opacity in the atmosphere of a magnetar.

Since the photon energies under consideration here are typically above 2 keV
(see \citealt{Woods2006a} for typical magnetar burst spectra), 
the opacity can be assumed to be dominated by Compton scattering off free electrons \citep{Thompson1995}, 
and photoelectric and atomic transition contributions are generally small.  Some corrections for these atomic processes are needed at slightly lower
photon energies, but these are not expected to significantly change our results.

The monochromatic opacity $\kappa_\nu$ follows from the Compton scattering cross section.
In a strong magnetic field, 
with its inherently anisotropizing influence,
this cross section depends strongly on the polarization mode of the photons.
Photons in the ordinary polarization mode (O-mode or $\parallel$ mode), with their 
electric field vector lying in the plane containing their direction
of propagation and the magnetic field, have a much larger chance of scattering than 
extraordinary mode (E-mode or $\perp$ mode) photons,
which are polarized perpendicular to this plane. The 
strong magnetic field also modifies the dielectric properties
of both the atmospheric/magnetospheric plasma, and polarizes
the vacuum \citep[][]{Adler1971a, Harding2006}.  
Accordingly, the field introduces profound modifications to the scattering opacity.

Assuming both photon modes are polarized perpendicular to their direction of propagation, the scattering cross
section for a photon in mode $j$ (where $j=1$ is the E-mode and $j=2$ the O-mode) traveling at an angle $\theta$
with respect to the magnetic field direction is given by \citep{Ho2003b}
\begin{align}
\sigma_j=\frac{\sigma_\rmn{Th}}{1+K_j^2}\biggl[ & \frac{1}{2}\frac{\omega^2(1-K_j\,\rmn{cos}\,\theta)^2}
	{(\omega-\omega_\rmn{C})^2+\Gamma_\rmn{e}^2/4} \nonumber\\
	& + \frac{1}{2}\frac{\omega^2(1+K_j\,\rmn{cos}\,\theta)^2}
	{(\omega+\omega_\rmn{C})^2} + K_j^2\,\rmn{sin}^2\theta \biggr],
\label{eq:crosssec}
\end{align}
with
\begin{equation}
\Gamma_\rmn{e}=\frac{2e^2\omega_C^2}{3m_\rmn{e}c^3},
\end{equation}
where
$\sigma_\rmn{Th}$ is the Thomson cross section,
$\omega$ is the angular frequency of the photon,
and $\omega_\rmn{C} = eB/m_ec$ is the electron cyclotron frequency.
$\Gamma_\rmn{e}$ is the classical natural linewidth of the cyclotron resonance,
and satisfies $\Gamma_\rmn{e}/\omega_\rmn{C} = 2\fsc\, (B/B_\rmn{cr})/3$ where 
$\fsc = e^2/\hbar c$ is the fine structure constant.
When $\omega_\rmn{C}$ becomes a sizable fraction of $m_ec^2/\hbar$, relativistic 
corrections to this linewidth become mandatory \citep[][]{Baring2005a}.  
However, in our calculations we use the Rosseland mean opacity
(except in Section \ref{sec:nonlte}, where we discuss what happens when we revise this approximation),
which will be detailed below,
which means the electron cyclotron resonance
gets smoothed out. The precise numerical value of the 
line width is therefore not very important,
unless the local field substantially exceeds $10^{15}$ G.
The term $K_j$ incorporates the influences of plasma and vacuum 
dispersion on the scattering cross section, and
is detailed in Appendix \ref{sec:appendix}, where we show how this form reduces
to the various commonly used approximations, as given by \citet{Herold1979} and \citet{Ventura1979b}.

For radiation frequencies below the cyclotron frequency,
this form for the cross section reduces to
\begin{equation}
\sigma_\rmn{E}\sim\frac{\omega^2}{\omega_\rmn{C}^2}\sigma_\rmn{Th}, \quad \sigma_\rmn{O}\sim\sigma_\rmn{Th},
\label{eq:crosssec_red}
\end{equation}
with $\omega$ the radiation frequency, $\omega_\rmn{C}$ the electron cyclotron frequency and $\sigma_\rmn{Th}$
the Thomson scattering cross-section. 
Thus, the scattering cross section to O-mode photons is roughly constant,
while the cross section to E-mode photons scales roughly as $T^2/B^2$
when $B\gg T$.  The fact that the scattering cross sections are so different for the two different polarization
modes means that the atmosphere will have two distinct photospheres: one for each photon mode. As the scattering
cross section is significantly larger for O-mode photons than for E-mode photons, the O-mode
photosphere will be located outside the E-mode photosphere.

We use the cross sections formulated by \citet{Ho2003b}, as these 
incorporate the quantum electrodynamical,
vacuum polarization and plasma dispersion effects, making them correct in the high-density
regions of the atmosphere, where the commonly-used scattering cross sections in 
the absence of dispersion from \citet{Herold1979} break down. However, these cross sections
do not incorporate Klein-Nishina reductions, which come into play
when the electron thermal energy approaches its rest mass energy,
i.e. $kT\sim m_ec^2$. This is also the regime where kinematic modifications to the opacity become
significant, i.e., when the motions of the electrons must be accounted for in 
treating photon transport.
Neither of these effects will be considered in the present work, as an opacity equation incorporating
all of the above effects is not currently available in the literature.
A side effect of this opacity treatment is that it is impossible to reproduce the results of \citet{Paczynski1986}
with our models by setting the magnetic field to zero, as in the nonmagnetic case the existence of extended
atmospheres depends on the Klein-Nishina corrections.

To form the total opacity, the scattering differential cross section 
in Eq.~(\ref{eq:crosssec}) is averaged over angles by multiplying it by sin $\theta$ 
before integrating over $\theta$, so as to model an isotropic
distribution of photons.  This neglects 
the inherent transport-induced anisotropy of photons in the surface layers \citep[][]{Ozel2001a}
that will introduce some 
quantitative (but not qualitative) modification to the outermost 
portions of our atmospheric profiles.

The scattering cross sections contain two resonances: the electron cyclotron resonance, and the vacuum
resonance. While these resonances are crucial for spectral modeling, we find that in our models they
are not of great significance, being mostly smoothed out to relatively minor features in the opacity
due to the fact that we use the Rosseland mean opacity. As we
assume the photons to be in thermal equilibrium with the gas throughout
the atmosphere in the equation for radiation pressure, it is reasonable to extend this assumption
and take the opacity at any point to be the Rosseland mean opacity at the local temperature, as given by
\begin{equation}
 \label{eq:rosseland}
 \frac{1}{\bar{\kappa}}=\left[\int_{0}^{\infty}\!\frac{1}{\kappa_{\nu}}\frac{\partial B_{\nu}(T)}
 {\partial T}\mathrm{d}\nu\right]\left[\int_{0}^{\infty}\frac{\partial B_{\nu}(T)}
 {\partial T}\mathrm{d}\nu\right]^{-1},
\end{equation}
where $\nu$ is the frequency of the radiation, $\kappa_\nu$
is the monochromatic opacity and $B_\nu$ is the Planck function.
This form applies when the gradients in temperature arise on much 
larger spatial scales than the gradients for opacity.  As a consequence of averages
being formed for inverse opacities, more tenuous and low opacity domains 
dominate the determination of the Rosseland mean, so that the various resonance
regions do not contribute much. We define the Rosseland mean opacity to E-mode photons
as $\bar\kappa_\rmn{E}$, and the Rosseland mean opacity to O-mode photons as
$\bar\kappa_\rmn{O}$. 
We will discuss the shortcomings of using the Rosseland mean opacity  in Section \ref{sec:nonlte}.

For the purpose of the structure of the atmosphere, the opacity sets the radiative energy transport through the atmosphere,
as described by Equation \eqref{eq:nabla_rad},
as well as the radiative force, which determines the critical luminosity given by Equation \eqref{eq:Lcr}.
For both energy transport and radiative force the relevant photon flux is the net flux, as in both cases the effect of
identical photons propagating in opposite directions cancel out. Therefore, we will define an opacity $\kappa_\rmn{net}$, as the
average opacity to the net outward flux of photons:
\begin{equation}
 \kappa_\rmn{net}=\frac{F_\rmn{O}}{F}\bar{\kappa}_\rmn{O}+\left(1-\frac{F_\rmn{O}}{F}\right)\bar{\kappa}_\rmn{E},
 \label{eq:kappanet}
\end{equation}
where $F$ is the net outward flux and
$F_\rmn{O}$ is the net outward flux in O-mode photons.
This represents an opacity that is appropriately 
weighted for the relative local fluxes of the two polarization modes.

In a scattering event, a photon has a certain probability of changing modes.
These probabilities can be calculated from the cross sections for scattering into one specific polarization
mode and for X-rays in magnetar atmospheres are given by
\citep{Ulmer1994a, Miller1995}
\begin{align}
P_{\rmn{E}\rightarrow\rmn{O}} =&\, 1/4,\nonumber\\
P_{\rmn{O}\rightarrow\rmn{E}} \simeq&\, \omega^2/\omega_\rmn{C}^2.
\label{eq:Pswitch}
\end{align}
These equations are approximate and only valid in the limit $\omega\ll\omega_\rmn{C}$, which
is true close to the star, at least up to a few kilometers above the surface; 
they also apply to the Thomson regime, i.e. provided $\omega\ll m_ec^2/\hbar$.

At large optical depth the net opacity will be given by the opacity to E-mode photons.
O-mode photons will effectively be stuck in this region, as they have
a much larger scattering cross section than E-mode photons \citep[][]{Miller1995}. 
This does not mean there are no
O-mode photons deep in the atmosphere, merely that they move inwards as much as outwards, not contributing
to the net flux, and can only diffuse effectively by converting to the E-mode.

At lower optical depth O-mode photons will start to contribute to the net flux. To quantify this contribution, we
consider the probability an O-mode photon at optical depth $\tau_\rmn{O}$ has of escaping
the atmosphere without converting to the E-mode.
This probability is roughly given by
\begin{equation}
 P_\rmn{esc}=(1-P_{\rmn{O}\rightarrow\rmn{E}})^{ \tau_\rmn{O}^2}\quad ,
 \label{eq:Pesc}
\end{equation}
in diffusive regimes.
Any O-mode photon that escapes in this way contributes to a net outward flux of O-mode photons at all points
between its origin (the point where it converted from the E-mode) and the outer edge of the atmosphere.

The probability given by Equation \eqref{eq:Pesc} is effectively zero in the region where the optical depth to E-mode photons
is greater than one, since the optical depth to O-mode photons is several orders of magnitude larger. Thus,
as long as the total photon population
is divided more or less equally over the two modes, which should be true due to detailed statistical balance,
we can safely
assume the entire net outward flux to be in E-mode photons at $\tau_\rmn{E}=1$.
Between this point and the outer edge of
the atmosphere, these E-mode photons will scatter once on average, so that the fraction of E-mode photons
that scatter in any interval $\rmn{d}\tau_\rmn{E}$ in the region $1>\tau_\rmn{E}>0$ is approximately equal to $\rmn{d}\tau_\rmn{E}$,
assuming most of the net flux will still be in E-mode photons.
The fraction of the net flux that is in O-mode photons at a point $\tau_\rmn{E}=\tau '$
in this region is then given by
\begin{equation}
 \label{eq:fracO}
 \frac{F_\rmn{O}}{F}= -\int_1^{\tau '}\!P_{\rmn{E}\rightarrow\rmn{O}}P_\rmn{esc}\,\mathrm{d}\tau_E,
\end{equation}
Note that this integral is constructed from inner radii to outer ones, so that 
$\tau' < 1$ and optical depth declines with radius.
This equation can be used to track the approximate fraction of the net outward
flux of photons in the O-mode, and thus calculate $\kappa_\rmn{net}$.

To test our method of determining
the fraction of net outward photons in the O-mode we compare this to
the analysis and Monte Carlo simulations performed by \citet{Miller1995}. The fraction of net outward
photons in the O-mode at $\tau=0$ as given by Equation \eqref{eq:fracO} should match the fraction of escaping
photons in the O-mode as calculated by \citet{Miller1995}. We have tested this for a range of constant values of
$\omega/\omega_\rmn{C}$ and find that the fraction of photons that escape the atmosphere in the O-mode
is roughly given by $0.22 \omega/\omega_\rmn{C}$. \citeauthor{Miller1995} performed Monte Carlo
simulations to calculate the radiative force divided by flux, rather than the fraction of photons escaping in the
O-mode. However, he also performed an order of magnitude calculation to predict the results of his simulations,
where he calculated the fraction of photons escaping in the O-mode, which
he then multiplied with $\sigma_\rmn{Th}/c$ to get the force divided by flux. In this order of magnitude
treatment he calculated that the fraction of photons escaping in the O-mode is given by
$0.1 \omega/\omega_\rmn{C}$,
which is a factor two lower than our estimate. However, he also found that his final prediction for force divided
by flux was a factor two lower than the results of his simulations. Thus, if we take the results of his Monte Carlo
simulations and calculate back to the fraction of photons escaping
in the O-mode by dividing by $\sigma_\rmn{Th}/c$,
we find that our predictions almost exactly match his simulations.
Note that while for this comparison we integrate Equation \eqref{eq:fracO} with a constant value of $\omega/\omega_\rmn{C}$ to match the approach taken by \citeauthor{Miller1995}, in our atmosphere models this equation is solved simultaneously with the stellar structure equations (Equations \ref{eq:dPdr}-\ref{eq:ddMdr}), using the radially variable values of $\omega/\omega_\rmn{C}$ and $\tau_\rmn{O}$ and thus obtaining the correct radial dependence of $F_\rmn{O}/F$.

\subsection{Boundary conditions and method}
\label{sec:method}
We solve six differential equations of stellar structure, for pressure, temperature, atmosphere mass and optical depth to
O-mode photons, E-mode photons and a combination of both, as given by Equations \eqref{eq:dPdr}, \eqref{eq:dTdr} and
\eqref{eq:ddMdr} and three versions of Equation \eqref{eq:dtaudr}: for the optical depth to E-mode photons, O-mode photons,
and the net flux of photons.
We do this by integrating these equations
from the stellar surface at $R_\star=10$~km to $R_\rmn{end}$, which we choose as the point where the density becomes so small that
the atmosphere has effectively ended, choosing this cutoff density as $10^{-15}$ g cm$^{-3}$. This is in principle an arbitrary value,
but we have verified that increasing this value to $10^{-12}$ g cm$^{-3}$ does not change the results at all, indicating that our chosen value
is likely even lower than necessary. Additionally, we find no change in our results from fixing $R_\rmn{end}$ at 1000 km, which
is well beyond the expected size of the atmosphere.

We solve Equation (\ref{eq:fracO}) for the fraction of the net flux contained in O-mode photons
from the point $\tau_\rmn{E}=1$ to $R_\rmn{end}$.
We integrate with a straightforward iterative shooting procedure, using an eighth order Runge-Kutta method with
embedded Dormand-Prince error estimation from \citet{Press2007a}. We start with guesses for all undetermined parameters
at the stellar surface, and iterate until all the outer boundary conditions are met.

Our input parameters are the magnetic field strength at the surface $B_\star$,
the density at the base of the atmosphere $\rho_\star$ and the luminosity as seen by an observer at infinity $L_\infty$.
Further boundary conditions are provided by
the assumption that the atmosphere has
ended at $R_\rmn{end}$, giving $\tau=\tau_\rmn{O}=\tau_\rmn{E}=0$ and $\Delta M=0$ at $r=R_\rmn{end}$. 
The final boundary condition is provided by the assumption that $F_\rmn{O}/F=0$ at $\tau_\rmn{E}=1$. 

We calculate a series of atmospheres with a density at the base of the atmosphere $\rho_\star$ in the range $10^{3}-10^{7}$~g~cm$^{-3}$, and luminosities from
$10^{36}$ to $10^{42}$~erg~s$^{-1}$, repeating
this for magnetic field strengths at the surface of the star of $10^{14}$ and $10^{15}$~G,
values chosen to describe a typical magnetar. The range of densities is chosen to encompass the total atmospheric mass from
\citet{Paczynski1986}, which is $\sim 2\times 10^{20}$~g for all models in that work,
and to span a sensible density range at the base of the atmosphere,
as $10^3$~g~cm$^{-3}$ is roughly the atmospheric density limit for a cold neutron star, while
$10^7$~g~cm$^{-3}$ is roughly where it becomes impossible for X-rays to propagate through the gas, due to the
plasma frequency becoming greater than the typical photon frequency.

Summarizing, our model consists of solving the differential equations of stellar structure \citep[][]{Thorne1977}
from the surface of a magnetar
out to a point where the density is so low that the atmosphere has effectively ended. 
The input parameters of our model are the surface magnetic field strength, surface temperature of the
star and the total mass of the atmosphere.
We make the following assumptions. 
\begin{itemize}
\item The magnetic field is purely radial, with the radial dependence of a dipole field right above the magnetic pole.
This also means there is no radial magnetic pressure component, so that magnetic pressure can be ignored.
\item The atmosphere consists of pure fully ionized helium.
\item The radiation field is in local thermodynamic equilibrium with the gas.
\item The opacity is dominated by Compton scattering, with the scattering cross sections as given by
\citet{Ho2003b}.
\item Relativistic effects in scattering that can potentially modify the opacity somewhat
for temperatures in excess of 50 keV are neglected.
\item For purposes of the structure equations, the relevant opacity is the opacity that belongs to the
net outward photon flux, as given by the sum of the Rosseland mean opacities to E-mode and O-mode
photons, weighted by the relative contribution of those two modes to the net outward flux.
\end{itemize}

\section{Results}
\label{sec:results}
Our results differ dramatically from those found for non-magnetic atmospheres by \citet{Paczynski1986}.
We find no stable atmospheres with photospheric height greater than about 10 meters.
Although the radius of the photosphere
increases as surface temperature is increased, it does so by a very small amount.
Additionally, the temperature
at the photosphere is higher rather than lower for models with larger photospheric radius,
because the increase in surface temperature required
to make the atmosphere expand more than compensates for any temperature decrease caused by having a
larger photospheric radius.
Our results also show that the photospheres for E- and O-mode photons are always close together, 
both spatially and in terms of temperature, despite
the large difference in the opacity to these two different modes.
An overview of input and output parameters for our computed hydrostatic atmospheres 
is given in Table \ref{tab:results}.

\begin{figure}
\begin{center}
\includegraphics[width=\linewidth]{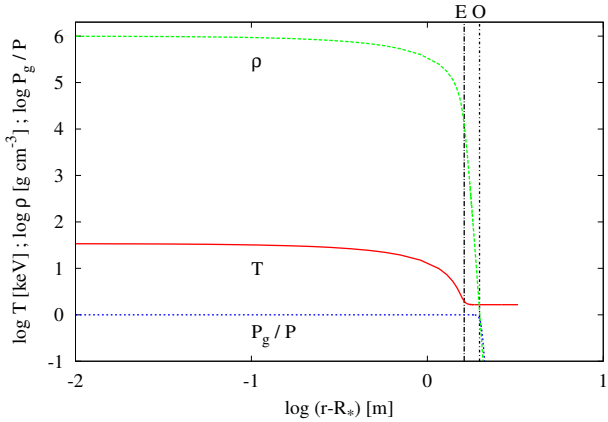}
\end{center}
\caption{
Radial structure of a magnetar atmosphere model, showing temperature $T$, density $\rho$
and gas pressure $P_\rmn{g}$  divided by total pressure $P$ plotted against radius for
$B=10^{14}$ G, $\rho_\star=10^{6}$ g cm$^{-3}$ and $L_\infty=10^{38}$~erg~s$^{-1}$. The two vertical lines represent
the E- and O-mode photospheres. Density and temperature drop rapidly below the E-mode photosphere. From there on
out, temperature stabilizes while density continues to drop, causing the total pressure to change from being gas pressure dominated
to radiation pressure dominated.
}
\label{fig:Tvsr}
\end{figure}

The general structure of the atmosphere in our results is  illustrated in Figure \ref{fig:Tvsr}.
Almost all the matter in the atmosphere is contained within the first ten meters, with only a fraction
$\sim 10^{-7}$ of the total atmospheric mass beyond that region. In those first few meters, the temperature drops rapidly
by about an order of magnitude, to stabilize from there on out in an unheated coronal region. 
The density rapidly drops down to the cutoff density where we define the atmosphere to end.
The pressure is dominated by gas pressure in the innermost
region, and becomes dominated by almost constant radiation pressure when the density drops. 
This structure is qualitatively the same for all different combinations of $L_\infty$, $B$
and $\rho_\star$, with different values giving only numerical differences.
In general, a higher luminosity gives a higher temperature throughout the atmosphere,
with luminosities that are too high causing the atmosphere to become unstable so that no solution to the equations can be found.
Higher magnetic field strengths mean a higher critical
luminosity due to a reduced opacity, so that a higher luminosity can propagate through the atmosphere before becoming unstable. Finally, a higher density gives higher temperatures and luminosities in optically thick regions for the same escaping luminosity, so that the luminosity limit at which the optically thick regions become unstable is lower.

The height of the photospheric radius of a few meters and the density structure of our model atmospheres
can be understood reasonably well in terms of the standard atmospheric scale height. When considered
at small height compared to the stellar radius, in an atmosphere composed of pure helium, above a 1.4 $M_\odot$
star, this scale height is given by
\begin{equation}
H \simeq 4 \frac{kT}{\rmn{[keV]}}\,\rmn{cm}.
\label{eq:scaleheight}
\end{equation}
Thus, for a typical model atmosphere with $kT_\star=25$ keV this scale height is about one meter. Over this length scale, the
density will drop by a factor $e$, which causes a strong drop in the temperature as well,
as the two gradients are linked in optically thick regions.
However, as the temperature drops, the scale height, which is proportional to temperature, decreases. This causes a rapidly
steepening gradient in the density and temperature profiles, so that the density decreases by many orders of magnitude inside
a few atmospheric scale heights, relative to that calculated at the surface. 
Over the range of this huge drop in density the material becomes optically
thin, which causes the temperature profile to cease tracing the mass density profile and signifies the photospheric radius.

\begin{figure}
\begin{center}
\includegraphics[width=\linewidth]{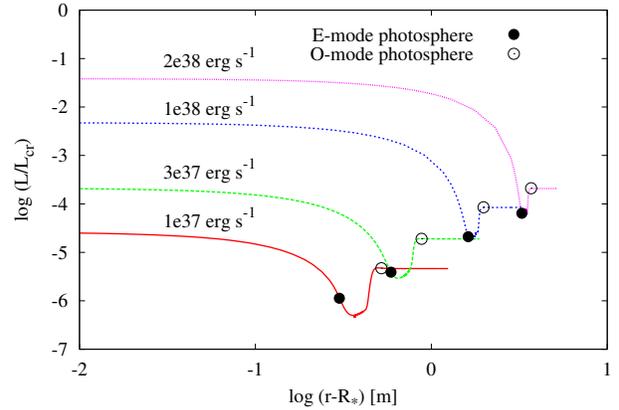}
\end{center}
\caption{
Stability of several model atmospheres, as given by the ratio between the local luminosity $L$ 
(which asymptotically approaches $L_{\infty}$ in 
Eq.~(\ref{eq:Stephan_Boltzmann_GR}) at large radii) and the critical luminosity
$L_\rmn{cr}$ plotted against the height above the stellar surface, for $B=10^{14}$ G,
$\rho_\star=10^6$~g~cm$^{-3}$ and $L_\infty=2\times10^{38},\,10^{38},\,3\times 10^{37}\,\rmn{and}\,10^{37}$~erg~s$^{-1}$, as marked in the figure.
The variations are mostly caused by the variations in opacity, with which $L/L_\rmn{cr}$ scales linearly, as
any other radius dependent terms are just gravitational corrections.
The initial downwards slope at low height is caused by the temperature dependence of the opacity through the photon
frequency. The sudden rise between the two photospheres is where the net radiation force switches from
being E-mode to O-mode dominated, causing an increase in opacity.
}
\label{fig:LLcr}
\end{figure}

Such gas pressure domination of the atmospheric structure close to the surface of the star
is also present in the nonmagnetic results from \citet{Paczynski1986}. However, in their
results a huge radiation pressure supported region, extending up to 200 km, sets up above this gas pressure supported region.
In contrast,
in our models, no significant radiation pressure supported region is present. The reason for this is that
for a radiation pressure supported region to be stable, the luminosity has to be almost equal to (but below) the critical luminosity
throughout this region \citep{Paczynski1986}. This is also the only way to have an extended atmosphere, as 
the thermal gas would need an unusually high (i.e., relativistic) temperature to support
a stable, extended atmosphere.
For an extended atmosphere to be in hydrostatic equilibrium requires
\begin{equation}
\frac{\rmn{d}P_\rmn{r}}{\rmn{d}r} = -\rho\frac{GM}{r^2}\left(1-\frac{R_\rmn{g}}{r}\right)^{-1/2}.
\end{equation}
Substituting for the gradient of radiation pressure using Equations (\ref{eq:dTdr},
\ref{eq:nabla_rad} and \ref{eq:pressure}) gives:
\begin{equation}
\frac{L}{L_\rmn{cr}}\left(1-\frac{R_\rmn{g}}{r}\right)^{-1/2} \simeq 1.
\label{eq:LequalsLcr}
\end{equation}
This equation is of general validity, and is satisfied in the nonmagnetic case 
for a large range of altitudes for an appropriate choice of temperature due to a fine balance
between the general relativistic and Klein-Nishina corrections \citep{Paczynski1986}.

\begin{figure}
\begin{center}
\includegraphics[width=\linewidth]{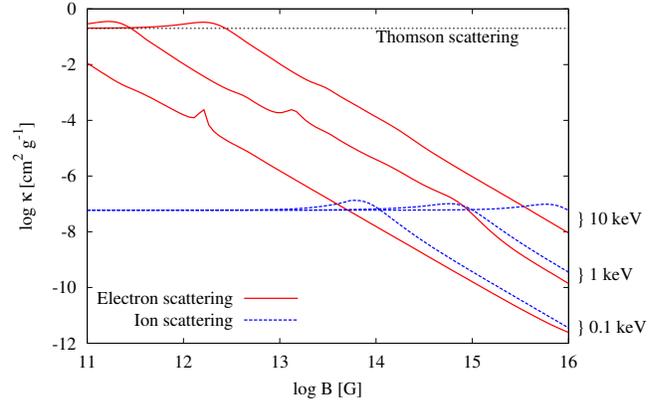}
\end{center}
\caption{
General behaviour of the electron and ion scattering opacities with differing temperature and magnetic field, illustrating
the dependencies on temperature and magnetic field strength of the opacity that determine the stability of our
atmosphere models.
Plotted are the Rosseland mean opacities for E-mode photons versus magnetic field strength,
showing both electron and ion scattering opacities,
 for $\rho=100$ g cm$^{-3}$
and $kT_\star=$10, 1 and 0.1 keV, as marked in the figure. These temperatures are chosen to show
the general temperature dependence. A different density would not change these curves as long as the gas
is still opaque to X-rays, except for the position of the vacuum resonance, which is the small resonance peak visible in
the electron scattering opacity around $10^{12}$ G for 0.1 keV and around $10^{13}$ G for 1 keV.
This resonance is present in all curves, but smoothed out to varying
degrees in calculating the Rosseland mean. The maximum of the curves lies at the cyclotron resonance,
and at lower magnetic field than this the opacity reduces to the Thomson opacity.
It is clear that the ion and electron scattering opacities are comparable in the high field limit, while the electron
scattering opacity dominates at lower fields, with the crossover between these two cases lying at higher magnetic field strength
for higher temperatures. In our models, the lowest temperatures are about 1 keV at $10^{14}$ G and about 3 keV at
$10^{15}$ G, so all our models lie in the region where electron scattering dominates, justifying our choice to neglect the
ion scattering opacities. Note that due to the nature of the Planck function, the peaks caused by the various resonances in the Rosseland mean opacity do not occur at the same position in these graphs as they would for the corresponding monochromatic opacities.
}
\label{fig:opacities}
\end{figure}

Figure \ref{fig:LLcr} shows the ratio between luminosity and critical luminosity in a magnetar atmosphere as it
follows from our models. This figure is analogous to
Figure 2 of \citet{Paczynski1986}, but while their results show that the luminosity
remains within a factor $10^{-4}$ below the critical luminosity up to  a height of 200 km,
our results show a fluctuation
of several orders of magnitude over that range, which shows that the equality given in Equation~(\ref{eq:LequalsLcr}) cannot be met
in our models.
These strong fluctuations in $L/L_\rmn{cr}$ are caused by the strong variations in opacity, which are illustrated in
Figure~\ref{fig:opacities}. This figure shows that the opacity in a magnetar atmosphere
drops rapidly with both decreasing temperature and increasing magnetic
field strength.

In our models, the temperature
drops rapidly up to the photosphere, so that the opacity also drops. The temperature only becomes close to constant
when it decouples from the density outside the photosphere, while for the photosphere to extend the opacity would have to
be close to constant over an extended region below the photosphere. Additionally, at much larger length scales the 
decreasing magnetic field strength would also prevent the opacity from being constant over an extended range of radii.

We thus find that extended hydrostatic magnetar atmospheres
in open field line regions do not exist, as the strong radial dependence of the opacity prevents the existence
of an extended radiation pressure supported region. This provides an important additional ingredient to the results
of our earlier work \citep{Watts2010}. While the criteria set out in that work for PRE to occur in magnetars
are all valid, its preliminary considerations omitted the requirement for near equality of the luminosity and the critical luminosity
throughout the atmosphere.  This turns out to be a criterion that cannot be met in open field line regions of
magnetars. In summation, the conditions given in \citet{Watts2010} are indeed necessary for PRE to occur, but they
are not sufficient.

\section{Maximum luminosity}
\label{sec:comparison}

We now consider the maximum luminosity that can propagate through the atmosphere of a magnetar
as it follows from our results, and compare this to previous work.
The critical luminosity in the atmosphere of a magnetar has been estimated several times in the literature.
Two results in particular are generally quoted. The first is the
estimate $L_\rmn{cr}=(\omega_\rmn{C}^2/\omega^2)L_\rmn{Edd}$ \citep{Paczynski1992a}, which
is based on the reduction of the scattering cross section of E-mode photons by the magnetic field
as compared to the nonmagnetic case. This estimate only takes into account E-mode photons, based on the
assumption that the entire luminosity will diffuse outwards in the form of E-mode photons, as these have a much
lower scattering cross section. The second estimate we compare our results with is the one
from \citet{Miller1995}, which
is based on the radiation force exerted by O-mode photons escaping the atmosphere,
and is given by $L_\rmn{cr}=5(\omega_\rmn{C}/\omega) L_\rmn{Edd}$. This equation is
based on a treatment in the region around the O-mode photosphere, which is where 
\citeauthor{Miller1995} expects the critical luminosity to be lowest. Therefore, we calculate this equation
at the O-mode photosphere, and treat the result as the maximum luminosity to be propagated through an
atmosphere without causing instability. Note that the critical luminosity in our models fully incorporates
the effects of gravitational redshift, while the given estimates do not, but this does not cause major
discrepancies.

\begin{figure}
\begin{center}
\includegraphics[width=\linewidth]{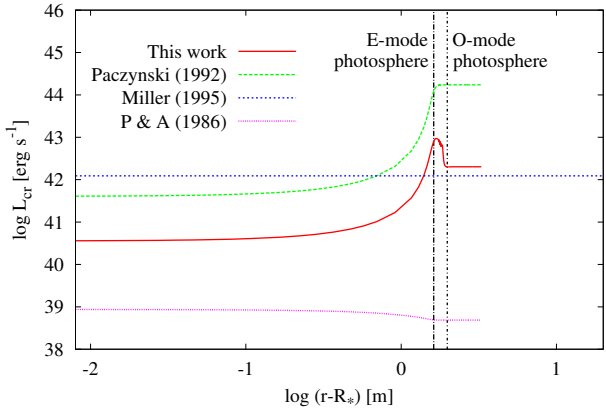}
\end{center}
\caption{Critical luminosity versus radius, comparing our numerical results to
 various theoretical predictions of this quantity, where the abbreviation P \& A refers to \citeauthor{Paczynski1986}. These results are for $B=10^{14}$ G,
$\rho_\star=10^{6}$~g~cm$^{-3}$ and $L_\infty=10^{38}$~erg~s$^{-1}$. Note that the Miller (1995) value is a straight line,
because Miller calculates the maximum luminosity that can be propagated through an atmosphere, rather than
a critical luminosity at each radius. The \citet{Paczynski1986} value is the nonmagnetic value.
}
\label{fig:luminosities}
\end{figure}

A comparison between these different critical luminosities is shown in Figure \ref{fig:luminosities},
and shows that the critical luminosity we find
is generally lower than previous predictions, although still significantly higher than non-magnetic limit.
Our results follow the behavior of the relation from
\citet{Paczynski1992a} up to the E-mode photosphere, but are an order of magnitude lower, which is caused
by the fact that we use the Rosseland mean opacity, which takes into account a thermal distribution of photons,
rather than the monochromatic opacity their estimate is based
on. Between the E-mode
and O-mode photospheres, our results switch to following the form from \citet{Miller1995}, as here the O-mode
photons start to dominate our opacity. 

Our results show that a single critical luminosity for the entire atmosphere,
or the maximum luminosity that can be propagated through the atmosphere, is roughly given by the lowest luminosity
out of two approximations. The first is the approximation in the E-mode dominated region following
from our models, where the critical
luminosity is the minimum value that $0.1(\omega_\rmn{C}/\omega)^2L_\rmn{Edd}$ takes on below the E-mode
photosphere. The second is the approximation for the region where O-mode photons dominate the radiation
force, which gives a critical luminosity of roughly $5\omega_\rmn{C}/\omega L_\rmn{Edd}$, calculated
at the O-mode photosphere, as given by \citet{Miller1995}. In our models the former is always the lower
of the two, but this does not necessarily have to be the case in more detailed atmosphere models.
These approximations break down when they get close to the nonmagnetic Eddington luminosity, where
the critical luminosity will become equal to this nonmagnetic value.

In our results, we find that this critical luminosity for the whole atmosphere is never approached closely. This is because
in our models this critical luminosity is set just above the surface
of the star, where the pressure is dominated by gas pressure. The critical luminosity gives the luminosity at which radiation
force balances gravitational force, but if there is significant gas pressure, this will also contribute to the force balance. Thus,
in a gas pressure dominated region the maximum luminosity that can be propagated without violating hydrostatic equilibrium
will actually be lower than the critical luminosity we calculate above, with the precise value depending on the atmospheric structure.

\section{Additional physics}
\label{sec:extensions}

Throughout this work we have made a number of simplifying assumptions. None of these assumptions
qualitatively alters our main result: extended hydrostatic atmospheres of magnetars cannot exist, because the near equality
of the luminosity and the critical luminosity is not possible over an extended range of radii. 

Three of our main simplifying assumptions are: assuming the validity of using the Rosseland mean opacity, ignoring the presence of an electron-positron pair plasma in the composition,
and ignoring magnetic confinement effects. We will now elaborate on why relaxing these assumptions would not
enable extended hydrostatic atmospheres of magnetars.

\subsection{Grey opacities}
\label{sec:nonlte}

Throughout our models, we have been using the Rosseland mean opacity as a way of reducing a
photon energy dependent problem to a simpler grey atmosphere problem. However, this has two main limitations.
Firstly, the Rosseland mean opacity is generally only a good approximation deep in the atmosphere, up to somewhere a little below the E-mode photosphere. Secondly,
due to the fact that the Rosseland mean opacity emphasizes the low-opacity parts of the spectrum, the outward radiation force is underestimated, particularly in the area around the vacuum resonance.

The first effect of using the Rosseland mean opacity in the outer regions of the atmosphere is that we find an incorrect temperature
profile in that part of the atmosphere. This effect has been quantified by \citet{Ozel2001a}, who showed that while a Rosseland mean opacity model gives a constant temperature profile at low optical depth, a full radiative transfer approach gives a temperature that continues to decline. 

The effect that  a more accurate temperature profile would have on our models is fairly straightforward to predict. The optically thin regions would still be dominated by radiation pressure, as the gas pressure falls away  rapidly. The temperature of the O-mode photosphere would be somewhat lower, and thus the value at which the critical luminosity stabilizes in the outer regions of the atmosphere would be higher by the same factor, since beyond the O-mode photosphere critical luminosity is set by the temperature of the O-mode photosphere, following the scaling of $\omega^{-1}$ predicted by \citet{Miller1995}. The critical luminosity would still stabilize at a constant value, as the photons do not thermalize with the gas in this region, so the continuing decrease of the temperature would have little effect on the radial structure of the atmosphere. Thus, a more accurate temperature profile would change our quantitative results slightly, but not our qualitative results.

The second problem with using the Rosseland mean opacity is more difficult to quantify. The Rosseland mean
opacity is the correct energy averaged opacity for energy transport, as it is used in Equation \eqref{eq:nabla_rad}. However, for the purposes
of the radiation force, the correct energy averaged opacity is the flux mean opacity, which is only equivalent to the Rosseland mean in thermodynamic equilibrium. Although there is no explicit radiation force
term in our equations, the form given for the radiation pressure in Equation \eqref{eq:pressure} is based on the same assumption of local thermodynamic equilibrium as the Rosseland mean opacity. So to correct our treatment of the radiation force in the outer regions of our atmosphere models, where local thermodynamic equilibrium is a very poor assumption, we have to calculate the radiation pressure from the radiation flux and the flux mean opacity through
\begin{equation}
\frac{\rmn{d}P_\rmn{r}}{\rmn{d}r}=-\frac{\kappa_\rmn{F}\rho L_\infty}{4\pi r^2 c}\left(1-\frac{R_\rmn{g}}{r}\right)^{-1},
\label{eq:dPrdr}
\end{equation}
where $\kappa_\rmn{F}$ is the flux mean opacity, which is defined as
\begin{equation}
\kappa_\rmn{F} = \left[\int_0^\infty\!\kappa_\nu F_\nu\,\rmn{d}\nu\right]\left[\int_0^\infty\! F_\nu\,\rmn{d}\nu\right]^{-1},
\label{eq:kappaflux}
\end{equation}
where $F_\nu$ is the monochromatic flux.
Equation \eqref{eq:dPrdr} adds an additional differential equation to our model, for which a boundary condition can be provided by requiring $P_\rmn{r}=F/c$ at the outer boundary atmosphere, where $F$ is the total flux. Additionally, the flux mean opacity should be used in Equation \eqref{eq:Lcr} instead of the Rosseland mean opacity, as this equation is based on equating the forces of radiation and gravity. In both these cases we still have to calculate $\kappa_\rmn{net}$ as defined in Equation \eqref{eq:kappanet}, but from the single mode flux mean opacities rather than from the single mode Rosseland mean opacities.

While ideally we would use the flux mean opacity as outlined above, calculating the flux mean opacity requires detailed radiative transfer equations. However, we can use different grey opacities, which can be approximated without radiative transfer modeling, to get an idea of the qualitative effect that using the flux mean opacity would have on our models. 

The Rosseland mean opacity will generally be lower than the flux mean opacity, as it strongly emphasizes the low-opacity part of the photon spectrum. The flux mean opacity also emphasizes the low-opacity energies somewhat, as the energy distribution of the flux will be skewed towards those energies were opacity is lower, but not as strongly as the Rosseland mean opacity. Thus, we can treat our models using the Rosseland mean opacity as a sort of lower boundary to a more accurate result, in which the radiation force is underestimated.

\begin{figure}
\begin{center}
\includegraphics[width=\linewidth]{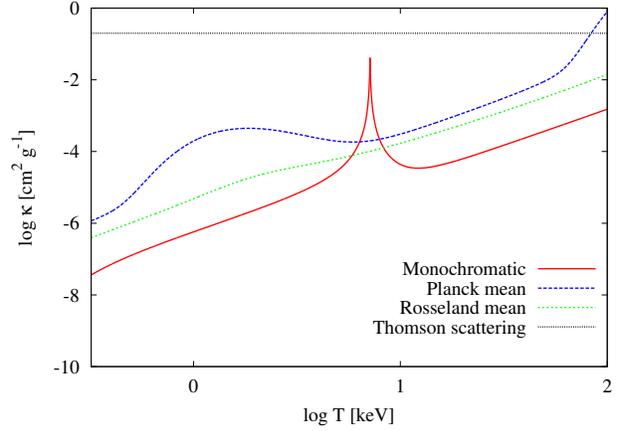}
\end{center}
\caption{Thomson scattering opacity and monochromatic, Planck mean and Rosseland mean opacities versus temperature, all for the E-mode. This is for magnetic field strength $B=10^{14}$ G and density $\rho=100$ g cm$^{-3}$, which is typical for the density around the vacuum resonance in our models. The peaks between 1 and 10 keV are cause by the vacuum resonance, while the sharp rise in the Planck mean opacity near 100 keV is caused by the electron cyclotron resonance. }
\label{fig:opacities_vsT}
\end{figure}

Similarly, we can create a set of models that function as a sort of upper boundary, where radiation force is overestimated, by using the Planck mean opacity. This opacity is defined as
\begin{equation}
 \label{eq:planck}
 \kappa_\rmn{P}=\left[ \int_0^\infty\!\kappa_\nu B_\nu \rmn{d}\nu \right] \left[\int_0^\infty\! B_\nu\rmn{d}\nu\right]^{-1}.
\end{equation}
In a region where the frequency-dependent opacity $\kappa_\nu$ changes slowly compared to the photon mean free path, the Planck mean opacity will be higher than the flux mean, as the Planck mean opacity does not emphasize any part of the photon spectrum, while the flux will be skewed to low-opacity energies. Since the temperature gradient in our models is set by photon diffusion, this should be roughly valid throughout our models, so that it is difficult to imagine the Planck mean opacity being lower than the flux mean opacity anywhere. Note that while the opacity to the net outward flux of photons $\kappa_\rmn{net}$ does change rapidly between the E- and O-mode photospheres, this is unrelated to the grey opacity used, but an effect of the mode switching of photons. Thus, it is reasonable to assume that the Planck mean opacity will be either larger than or comparable to the flux mean opacity throughout our models, and thus provide an interesting upper boundary to the effect of radiation force on our models.
The difference between the monochromatic, Rosseland mean and Planck mean opacities is illustrated in Figure \ref{fig:opacities_vsT}.

An additional argument for using the Planck mean opacity can be made by looking at radiative transfer calculations of magnetar emission spectra. Such calculations \citep{Ozel2001a, Ho2001a, van-Adelsberg2006a, Suleimanov2009a} typically show emerging spectra that are very similar to blackbody, with large deviations only occurring due to the proton cyclotron resonance. This resonance is not included in our models, but falls inside the photospheres for some of our models with $B=10^{15}$ G, where it occurs at roughly 6 keV. However, as the frequency at which the proton cyclotron resonance occurs is constant throughout the atmosphere, the photon flux at this resonance will be greatly reduced, so this resonance will have little influence on the flux mean opacity. Thus, the Planck mean opacity represents not only an interesting bound to the radiation force in our problem, but actually gives a reasonable approximation to the flux mean opacity.

Using the additional differential equation given by Equation \eqref{eq:dPrdr}, substituting the Planck mean opacity for the flux mean opacity, and replacing the Rosseland mean opacity by the Planck mean opacity in the equation for the critical luminosity (but not in Equation \eqref{eq:nabla_rad} where the Rosseland mean is the correct grey opacity) we construct a new set of model atmospheres. The results from these models are summarized in Table \ref{tab:nonlteresults}. The main difference between these results and our previous results is that the highest luminosities for which we find stable atmosphere solutions are even lower than the luminosities we found previously, generally by a factor of a  few.

\begin{figure}
\begin{center}
\includegraphics[width=\linewidth]{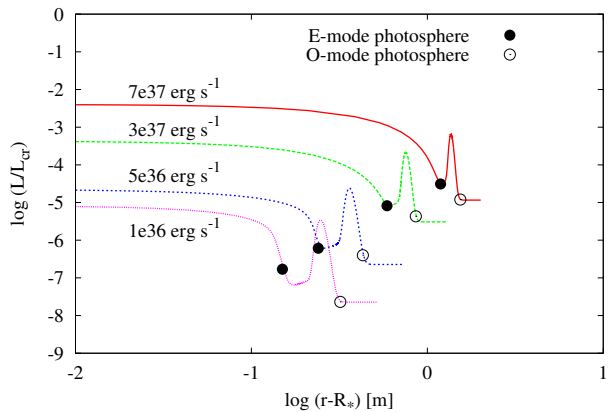}
\end{center}
\caption{
Local luminosity as a fraction of critical luminosity versus radius, for models using the Planck mean opacity, using different values of $L_\infty$ as indicated in the figure, for $B=10^{14}$ G and $\rho_\star=10^6$ g cm$^{-3}$. This figure is analogous to Figure \ref{fig:LLcr}, and the behaviour is similar to that observed in that figure. The exception is the region between the two photospheres, where the opacity, and thus the critical luminosity, becomes dominated by the vacuum resonance. This is what causes the sharp peak seen between the two photospheres. The values for $L_\infty$ include the highest value that gives a stable solution, and are otherwise chosen to aid visual clarity.  }
\label{fig:LLcr_17}
\end{figure}

The radial structure of the atmosphere in these models is practically identical to the structure detailed in Section \ref{sec:results} and shown in Figure \ref{fig:Tvsr}, namely a compact gas pressure supported atmosphere, with no significant contribution from radiation pressure. While the behaviour of the opacity just below the O-mode photosphere in these models is significantly different from that in our previous models due to the much larger influence of the vacuum resonance, the Planck mean opacity does not become (significantly) larger around the vacuum resonance than it is at the hot base of the atmosphere. This can also be seen in Figure \ref{fig:LLcr_17}, which shows the variation of the ratio between the luminosity and the critical luminosity with radius. While this figure shows that the detailed behaviour of the critical luminosity is different compared to our previous models, the strong radial dependence and the absence of a radiation pressure supported region remain.  Thus, the structure of the atmosphere in these models is relatively unaffected compared to our LTE models. Furthermore, the strong vacuum resonance in the Planck mean opacity introduces an additional radial dependence into the critical luminosity, strengthening our qualitative argument as to the impossibility of an extended radiation pressure supported region.

\subsection{Electron-positron pair plasma}
\label{sec:composition}
For the composition we have assumed an atmosphere composed entirely of fully ionized
Helium. Changing this to a different atomic composition would
change the results by a small numerical factor, and thus not have any significant impact on our conclusions.
However, a realistic magnetar atmosphere will also contain an electron-positron plasma,
particularly if it is hot enough to contain mildly-relativistic electrons.
This could make a significant difference, as it adds scattering particles
without adding a significant amount of mass, thus increasing the opacity.

The number density of a one-dimensional, magnetized, electron-positron plasma is given by 
\citep{Canuto1977a, Thompson1995}
\begin{equation}
n_{\rmn{e}\pm} \; =\; \frac{B}{B_{cr}} \left( \frac{m_\rmn{e}c}{\hbar} \right)^3
    \left(\frac{kT }{2\pi^3\, m_\rmn{e}c^2}\right)^{1/2}e^{-m_\rmn{e}c^2/kT}.
\end{equation}
The appearance of the crucial exponential factor encapsulates the rest mass 
contribution to the pair chemical potential when the pairs are in equilibrium with photons.
Calculating this number density for some of our atmosphere models shows that the exponential dependence
on $T$ causes a drop of many orders of magnitude in the pair
plasma number density in the first few meters of the atmosphere,
where the temperature drops rapidly. This means that either the contribution of the 
pair plasma to the total number density of electrons in the
rest of the atmosphere is negligible, or that the number density of electrons in the high temperature region
at the base of the atmosphere
is incredibly large, which would then cause the critical luminosity to drop below the 
radiation luminosity.

The rapid temperature drop in the first few meters of the atmosphere
is also a feature of the models from \citet{Paczynski1986}, and 
likely of any magnetar atmosphere model. Thus,
we conclude that even including an electron-positron plasma 
in a hydrostatic model would not enable an extended atmosphere, although such
an extension is anticipated to influence the detailed atmospheric structure.

\subsection{Magnetic confinement}
The other major assumption in our model that we know to be wrong is treating the magnetic field as purely
radial, which means we ignore any magnetic confinement effects. As we noted in Section \ref{sec:Lcr},
the confinement caused by an 
closed field line region is too strong to allow PRE, as the required expansion will
never occur.

In a poloidal field structure, the non-radial (i.e., confining) component of the 
field scales as $\sin\Theta$ at colatitudes $\Theta$.
Even if such a small
closed field component were to contribute to the confinement, Equation \eqref{eq:intro:Lcr_mag}
has the same $B^2$ dependence
as the critical luminosity from Compton scattering. The same can be asserted for 
quasi-toroidal field components from non-poloidal field morphologies.  Accordingly, the vertical 
gradient of the non-radial component of the magnetic field would be large.
Thus, just as for Compton scattering, magnetic confinement
cannot create near-equality of the luminosity and critical luminosity. 
The only way 
for the field stress to
keep the critical luminosity roughly constant, and thus equal to the luminosity, over an extended range of radii,
would be for the closed field component to be at most a very weakly-dependent 
function of radius.  This is highly unlikely in any
reasonable magnetic field geometry.

\section{Conclusions}
\label{sec:conclusions}

Hydrostatic extended magnetar atmospheres do not exist, due to the fact that
the strong dependence of opacity on radius makes the required near equality of the luminosity and the
critical luminosity throughout the atmosphere impossible. 
A hypothetical extended magnetar atmosphere will thus either have large regions where luminosity is greater
than the critical luminosity, or it will have large regions where the luminosity is much lower than the
critical luminosity, so that it is not supported against gravity. Therefore, such an atmosphere
cannot exist. This is unlike the nonmagnetic case, where a precise balance between luminosity and critical
luminosity means the photospheric radius can extend up to 200 km. The fact that magnetars do not have extended
hydrostatic atmospheres means that PRE, as envisaged previously, cannot work, due to the fact that stable expansion
of the atmosphere is not possible.

Additionally, we find that the maximum luminosity
that can be propagated through a hydrostatic magnetar atmosphere may be lower than the critical
luminosity given by \citet{Miller1995}, depending on the structure of the atmosphere. This is due to the fact that
\citet{Miller1995} assumes the maximum luminosity that can propagate will be set by the scattering of O-mode
photons near the O-mode photosphere. However, in our models the maximum luminosity is set by
E-mode photons scattering in the highest temperature region near the surface of the star, where the scattering cross
section is relatively large due to the high frequency of the thermalized photons. This means that depending on the
atmospheric structure, more observed magnetar bursts might have reached their critical luminosity than previously
assumed.

Our results have implications for interpretation of spectral fits to magnetar burst data. Magnetar bursts are typically
fit with several different spectral models, the main ones being two power laws, a power law plus a black body, two black
bodies, and more recently a power law with exponential cutoff and optically thin thermal bremsstrahlung.
The two black body model is typically one of the best fitting of these models \citep{Olive2004a, Feroci2004a},
with typical fitting parameters giving two temperatures around 3 and 11 keV, and typical emission region radius of roughly
the radius of a neutron star for the colder black body, and an order of magnitude smaller than that for the hotter component,
but with a large scatter in the sizes.  While this model has mostly been
presented as purely phenomenological, it has occasionally been suggested that this could be
interpreted as representing the
distinct signatures of the E- and O-mode photospheres \citep[][]{Israel2008a, Kumar2010a}.
This interpretation was attractive in the sense that it opened up the possibility of another way
of measuring the critical luminosity.
Our results show that in a hydrostatic atmosphere the temperature difference between these
two photospheres is never more than one or two keV, and that they are very close to
each other spatially. This makes the interpretation of the two blackbody model as representing the 
E- and O-mode photospheres highly unlikely.

Our results pose the question of what does happen when a magnetar burst exceeds the critical luminosity.
As we have shown, hydrostatic extended atmospheres are impossible. Thus, solving this problem
will require dynamical, time dependent models. These models would very likely result
in outflows, which could give rise to several forms of observable emission, such as the radio emission
seen from the outflow from the SGR 1806-20 Giant Flare \citep{Cameron2005a, Gaensler2005, Granot2006}, or
the `Magnetar Wind Nebula' recently detected around Swift J1834.9-0846 \citep{Younes2012a}.
Any outflows would
likely be very sudden, due to the short sound crossing time and burst time scale involved.
The observation that prompted this research, that of SGR J0501+4516,  likely reached
or exceeded its critical luminosity 
\citep{Watts2010}, and would thus be a prime candidate for testing any future dynamical models of magnetar
bursts.

\section*{Acknowledgments}
The authors would like to thank Cole Miller, Wynn Ho, Stuart Sim and Ralph Wijers for useful discussions.  TvP acknowledges support from the ERC through Advanced Grant no. 247295. ALW and CD'A acknowledge support from an NWO Vidi grant (PI Anna Watts).   MGB thanks the NASA Astrophysics Theory Program for support through grant NNX10AC59A.  CK is partially supported by NASA grant NNH07ZDA001-GLAST.


\appendix

\section{Details of the scattering cross section equation}
\label{sec:appendix}
This appendix gives more detail on the term $K_j$ from Equation \eqref{eq:crosssec}, 
which incorporates the effects of vacuum polarization and plasma dispersion into the Compton scattering
cross sections for photons in a super-strong magnetic field. The index $j$ indicates the polarization mode, where
$j=1$ stands for the E-mode and $j=2$ for the O-mode. 
This term is what differentiates the treatment by \citet{Ho2003b} from older commonly used approximations
\citep{Herold1979, Ventura1979b}.
We will now detail what $K_J$ is with and without including vacuum polarization and plasma dispersion effects.
$K_j$ is given by
\begin{equation}
K_j=\beta\left[1+(-1)^j\left(1+\frac{1}{\beta^2}\right)^{1/2}\right],
\end{equation}
\begin{equation}
\beta=\beta_0\beta_V,
\end{equation}
\begin{equation}
\beta_0 = \frac{\omega_\rmn{C}}{\omega}\frac{1}{2(1-v_\rmn{e})}
	\frac{\rmn{sin}^2\theta}{\rmn{cos}\,\theta},
\end{equation}
\begin{equation}
v_\rmn{e} = \frac{\omega_\rmn{P}^2}{\omega^2},
\end{equation}
where $\omega_\rmn{P}$ is the plasma frequency.
The parameter $\beta_V$ describes the influence of vacuum polarization. If vacuum polarization can be neglected,
$\beta_V=1$, and the cross section reduces to the form given by \citet{Ventura1979b}, which includes plasma dispersion
and quantum electrodynamical effects, but not vacuum polarization. 
If plasma dispersion is also neglected, $\beta_0\gg 1$, so that $K_1\sim 0$ and $K_2 \rightarrow \infty$, in which limit
Equation \eqref{eq:crosssec} reduces to the form given by \citet{Herold1979}.

When including vacuum polarization, the parameter $\beta_V$
is given by \citep{Ho2003b}
\begin{align}
\beta_V = & \left(1+\frac{\hat{a}+q}{1-v_\rmn{e}}\right)^{-1}\Biggl[1+\frac{(q+m)(1-u_\rmn{e})}
	{u_\rmn{e}v_\rmn{e}} \biggl(1-\frac{\hat{a}+m}{q+m}v_\rmn{e} \nonumber\\
	& -\frac{v_\rmn{e}(1-Mu_\rmn{i})}{1-u_\rmn{e}}
	\frac{-\hat{a}+q+m(1-v_\rmn{e})}{q+m}\biggr) \Biggr],
\end{align}
\begin{align}
\hat{a} \simeq & \frac{\alpha_\rmn{F}}{2\pi}\left[1.195-\frac{2}{3}\rmn{ln}\,\frac{B}{B_\rmn{cr}}-\frac{B_\rmn{cr}}{B}
	\left(0.8553+\rmn{ln}\,\frac{B}{B_\rmn{cr}}\right) -\frac{B_\rmn{cr}^2}{2B^2}\right], \nonumber\\
q \simeq & -\frac{\alpha_\rmn{F}}{2\pi} \biggl [-\frac{2}{3}\frac{B}{B_\rmn{cr}} +1.272-\frac{B_\rmn{cr}}{B}
	\left(0.3070+\rmn{ln}\,\frac{B}{B_\rmn{cr}}\right) & \nonumber\\
	& \hspace{1cm} -0.7003\frac{B_\rmn{cr}^2}{B^2}\biggr], \nonumber\\
m \simeq & -\frac{\alpha_\rmn{F}}{2\pi} \left[\frac{2}{3} + \frac{B_\rmn{cr}}{B}
	\left(0.1447-\rmn{ln}\,\frac{B}{B_\rmn{cr}}\right) -\frac{B_\rmn{cr}^2}{B^2}\right],
\end{align}
\begin{equation}
u_\rmn{e} = \frac{\omega_\rmn{C}^2}{\omega^2}, \quad 
u_\rmn{i} = \frac{\omega_\rmn{C,ion}^2}{\omega^2},
\end{equation}
where
$\omega_\rmn{C,ion}$ is the ion cyclotron frequency,
$\alpha_\rmn{F}$ is the fine-structure constant and
$M=\omega_\rmn{C}/\omega_\rmn{C,ion}$.
These equations have been derived under the assumption $u_\rmn{i}\ll1$, which is consistent with our results,
as well as $v_\rmn{e}\leq1$, which has to be true for radiation to be able to propagate.

\clearpage

\begin{deluxetable}{ccc cccccc}
\tabletypesize{\small}
\tablewidth{0pt}
\tablecaption{
Properties of the computed magnetar atmospheres using the method described in Section \ref{sec:model}. Columns show the magnetic field strength, density at the base of the atmosphere, luminosity
surface temperature, total atmosphere mass, radius and temperature of the E-mode
photosphere and radius and temperature of the O-mode photosphere. For each combination of magnetic field strength and surface density we show the result for the highest luminosity that still gives a stable atmosphere, and the two round powers of ten in luminosity below that. A value of 0 for the height of the E-mode photosphere means that that particular atmosphere model is optically thin to E-mode photons.
}
\tablehead{ \multicolumn{3}{c}{Input variables} &&& \multicolumn{2}{c}{E-mode photosphere} 
         & \multicolumn{2}{c}{O-mode photosphere}\\
	\colhead{$B_\star$} & \colhead{$\rho_\rmn{b}$} & \colhead{$L_\infty$} & \colhead{$kT_\star$}
	& \colhead{$\Delta M_\rmn{tot}$} & \colhead{$R-R_\star$} & \colhead{$kT$}
	& \colhead{$R-R_\star$} & \colhead{$kT$}  \\
	(G) & (g cm$^{-3}$) & (erg s$^{-1}$) & (keV) & (g) & (m) & (keV) & (m) & (keV) \\ }

\tablecolumns{10}
\startdata
$10^{14}$ & $10^{4}$ & $1 \times 10^{38}$   & 1.9 & $1.3 \times 10^{17}$ &  0 	&  1.9	&  0.4	& 1.7  \\
$10^{14}$ & $10^{4}$ & $1 \times 10^{39}$   & 6.3 & $2.4 \times 10^{18}$ &  0.3	&  3.5	&  0.8	& 3.0  \\
$10^{14}$ & $10^{4}$ & $4 \times 10^{39}$   & 24 & $9.8 \times 10^{18}$ &  1.2	&  4.9	&  1.9	& 4.2  \\
$10^{14}$ & $10^{5}$ & $1 \times 10^{37}$   & 1.4 & $5.2 \times 10^{18}$ &  0.05	&  1.1	&  0.3	& 1.0 \\
$10^{14}$ & $10^{5}$ & $1 \times 10^{38}$   & 4.7 & $1.8 \times 10^{19}$ &  0.2	&  2.0	&  0.6	&  1.7 \\
$10^{14}$ & $10^{5}$ & $6 \times 10^{38}$   & 24 & $9.1 \times 10^{19}$ & 1.2 	&  3.1	&  1.7	& 2.6  \\
$10^{14}$ & $10^{6}$ & $1 \times 10^{37}$   & 6.3 & $2.3 \times 10^{20}$ & 0.3 	&  1.1	&  0.5	& 0.9  \\
$10^{14}$ & $10^{6}$ & $1 \times 10^{38}$   & 34 & $1.3 \times 10^{21}$ & 1.6 	&  1.9	&  2.0	& 1.6  \\
$10^{14}$ & $10^{6}$ & $2 \times 10^{38}$   & 68 & $2.6 \times 10^{21}$ & 3.3 	&  2.3	&  3.7	& 2.0  \\
\\
$10^{15}$ & $10^{4}$ & $1 \times 10^{40}$   & 6.0 & $2.2 \times 10^{18}$ &  0	& 6.0 	&  1.4	& 5.8 \\
$10^{15}$ & $10^{4}$ & $1 \times 10^{41}$   & 10.9 & $4.1 \times 10^{18}$ & 0.02 	& 10.8 	&  2.4	& 9.1 \\
$10^{15}$ & $10^{4}$ & $6 \times 10^{41}$   & 25 & $9.9 \times 10^{18}$ &  0.7	& 17	&8.2  	& 15  \\
$10^{15}$ & $10^{5}$ & $1 \times 10^{39}$   & 3.4  & $1.3 \times 10^{19}$ & 0 	&  3.4	&  	0.9& 3.1  \\
$10^{15}$ & $10^{5}$ & $1 \times 10^{40}$   & 9.7 & $3.6 \times 10^{19}$ & 0.3 	& 6.2 	&  1.7	& 5.3  \\
$10^{15}$ & $10^{5}$ & $5 \times 10^{40}$   & 27 & $1.0 \times 10^{20}$ & 1.2 	&  9.2	&  	3.2& 7.9  \\
$10^{15}$ & $10^{6}$ & $1 \times 10^{39}$   & 7.1 & $2.6 \times 10^{20}$ & 0.3 	& 3.5	& 1.2 	& 3.0  \\
$10^{15}$ & $10^{6}$ & $1 \times 10^{40}$   & 60 & $2.3 \times 10^{21}$ & 2.9 	&  6.2	&  4.3	& 5.3  \\
$10^{15}$ & $10^{6}$ & $4 \times 10^{40}$   & 168 & $8.1 \times 10^{21}$ & 9.4 	&  8.7	&  11.4	& 7.5  \\

\enddata

\label{tab:results}
\end{deluxetable}

\begin{deluxetable}{ccr cccccc}
\tabletypesize{\small}
\tablewidth{0pt}
\tablecaption{
Properties of the computed magnetar atmospheres using the non-LTE method described in Section \ref{sec:nonlte}. 
}
\tablehead{ \multicolumn{3}{c}{Input variables} &&& \multicolumn{2}{c}{E-mode photosphere} 
         & \multicolumn{2}{c}{O-mode photosphere}\\
	\colhead{$B_\star$} & \colhead{$\rho_\rmn{b}$} & \colhead{$L_\infty$} & \colhead{$kT_\star$}
	& \colhead{$\Delta M_\rmn{tot}$} & \colhead{$R-R_\star$} & \colhead{$kT$}
	& \colhead{$R-R_\star$} & \colhead{$kT$}  \\
	(G) & (g cm$^{-3}$) & (erg s$^{-1}$) & (keV) & (g) & (m) & (keV) & (m) & (keV) \\ }

\tablecolumns{10}
\startdata
$10^{14}$ & $10^{4}$ & $1 \times 10^{38}$   & 1.9 & $7.1 \times 10^{17}$ & 0 	&  1.9	&  0.4	& 1.6  \\
$10^{14}$ & $10^{4}$ & $1 \times 10^{39}$   & 6.4 & $2.4 \times 10^{18}$ & 0.3 	&  3.4	&  0.8	& 3.0  \\
$10^{14}$ & $10^{4}$ & $3 \times 10^{39}$   & 20 & $8.9 \times 10^{18}$ &  1.1	&  4.6	&  1.7	& 3.9  \\
$10^{14}$ & $10^{5}$ & $1 \times 10^{37}$   & 1.4 & $5.2 \times 10^{18}$ &  0.06	&  1.1	&  0.2	& 0.9 \\
$10^{14}$ & $10^{5}$ & $1 \times 10^{38}$   & 4.7 & $1.8 \times 10^{19}$ & 0.2 	&  1.9	&  0.6	& 1.7  \\
$10^{14}$ & $10^{5}$ & $6 \times 10^{38}$   & 24 & $9.3 \times 10^{19}$ & 1.2 	&  3.0	&  1.7	& 2.5  \\
$10^{14}$ & $10^{6}$ & $1 \times 10^{37}$   & 6.3 & $2.3 \times 10^{20}$ & 0.3 	&  1.1	&  0.5	& 0.9  \\
$10^{14}$ & $10^{6}$ & $7 \times 10^{37}$   & 25 & $9.3 \times 10^{20}$ &  1.2	&  1.8	&  1.5	& 1.5	  \\
\\
$10^{15}$ & $10^{4}$ & $1 \times 10^{39}$	& 3.4	& $1.3 \times 10^{18}$	& 0	& 3.4	& 0.8	& 3.3 \\
$10^{15}$ & $10^{4}$ & $1 \times 10^{40}$	& 6.0	& $2.3 \times 10^{18}$	& 0	& 6.0	& 1.4	& 5.8 \\
$10^{15}$ & $10^{4}$ & $5 \times 10^{40}$	& 9.0	& $3.4 \times 10^{18}$	& 0	& 9.0	& 1.9	& 6.5 \\
$10^{15}$ & $10^{5}$ & $1 \times 10^{38}$	& 1.9	& $7.1 \times 10^{18}$	& 0	& 1.9	& 0.5	& 1.8 \\
$10^{15}$ & $10^{5}$ & $1 \times 10^{39}$	& 3.4	& $1.3 \times 10^{19}$	& 0	& 3.4	& 0.8	&  2.8\\
$10^{15}$ & $10^{6}$ & $1 \times 10^{37}$	& 1.9	& $6.8 \times 10^{19}$	& 0.05	& 1.1	& 0.3	&  1.0 \\
$10^{15}$ & $10^{6}$ & $1 \times 10^{38}$	& 3.0	& $1.1 \times 10^{20}$	& 0.1	& 1.9	& 0.6	&  1.6 \\

\enddata

\label{tab:nonlteresults}
\end{deluxetable} 

\label{lastpage}
\end{document}